\documentclass[aps,prl,twocolumn,groupedaddress]{revtex4}

\usepackage{longtable}
\usepackage{graphicx}
\usepackage{dcolumn}
\usepackage{bm}
\usepackage{amsmath}
\usepackage[psamsfonts]{amssymb}
\usepackage{subfigure}

\newcommand{\STO}{SrTiO$_3$}
\newcommand{\LAO}{LaAlO$_3$}

\newcommand{\etal}{\emph{et al.}}

\newcommand{\Rs}{R$_\square$}

\begin{document}

\title{Strong correlations elucidate the electronic structure and phase-diagram of \LAO/\STO\ interface}

\author{E. Maniv}
\affiliation{Raymond and Beverly Sackler School of Physics and Astronomy, Tel-Aviv University, Tel Aviv, 69978, Israel}
\author{M. Ben Shalom}
\affiliation{Raymond and Beverly Sackler School of Physics and Astronomy, Tel-Aviv University, Tel Aviv, 69978, Israel}
\author{A. Ron}
\affiliation{Raymond and Beverly Sackler School of Physics and Astronomy, Tel-Aviv University, Tel Aviv, 69978, Israel}
\author{M. Mograbi}
\affiliation{Raymond and Beverly Sackler School of Physics and Astronomy, Tel-Aviv University, Tel Aviv, 69978, Israel}
\author{A. Palevski}
\affiliation{Raymond and Beverly Sackler School of Physics and Astronomy, Tel-Aviv University, Tel Aviv, 69978, Israel}
\author{M. Goldstein}
\affiliation{Raymond and Beverly Sackler School of Physics and Astronomy, Tel-Aviv University, Tel Aviv, 69978, Israel}
\author{Y. Dagan}
\email[]{yodagan@post.tau.ac.il} \affiliation{Raymond and Beverly Sackler School of Physics
and Astronomy, Tel-Aviv University, Tel Aviv, 69978, Israel}

\maketitle

The interface between the two band insulators \STO\ and \LAO\ unexpectedly has the properties of a two dimensional electron gas. It is even superconducting with a transition temperature, T$_\textrm{c}$, that can be tuned using gate bias V$_\textrm{g}$, which controls the number of electrons added or removed from the interface. The gate bias - temperature (V$_\textrm{g}$, T) phase diagram is characterized by a dome-shaped region where superconductivity occurs, i.e., T$_\textrm{c}$ has a non-monotonic dependence on V$_\textrm{g}$, similar to many unconventional superconductors. In this communication the frequency of the quantum resistance-oscillations versus inverse magnetic field is reported for various V$_\textrm{g}$. This frequency follows the same nonmonotonic behavior as T$_\textrm{c}$; similar trend is seen in the low field limit of the Hall coefficient. We theoretically show that electronic correlations result in a non-monotonic population of the mobile band, which can account for the experimental behavior of the normal transport properties and the superconducting dome.

\section {Introduction}
When \LAO\ is epitaxially grown on TiO$_2$-terminated $\{100\}$ \STO, conductivity appears starting from a \LAO\ thickness threshold of four unit cells \cite{ohtomo2004high, thiel2006tunable}. The transport properties as well as the superconducting ones are strongly dependent on gate bias \cite{thiel2006tunable, caviglia2008electric, bell2009dominant, shalom2010tuning}. Recently, conducting interface has also been observed in $\{110\}$ \STO/\LAO~ interfaces \cite{annadi2013anisotropic}. Gate bias can be either applied by biasing the back of the substrate relative to the conducting layer (back gate) or by applying the electric field across the thin dielectric \LAO\ layer (top gate) \cite{hosoda2013transistor}. Superconductivity at the interface is observed over a carrier density range of a few $10^{13}$ cm$^{-2}$ but is suppressed at higher densities. In bulk \STO\ superconductivity is known to depend on the number of charge carriers \cite{SchooleydiscoveryofSCinSTO, BehniaFSofSTO}, and it appears in one or two bands depending on carrier density \cite{BehniaonsetofTWOtypes}. In contrast with \STO/\LAO, in bulk \STO\ superconductivity extends over a broad region of charge densities. It has been proposed that two-dimensional fluctuations destroy superconductivity in the low carrier regime \cite{caviglia2008electric}; However, it is still difficult to understand why superconductivity disappears at a relatively low density on the overdoped side of the phase diagram of \STO/\LAO.

Previously we have found evidence for the existence of multiple bands in \STO/\LAO\ \cite{shalom2010shubnikov}. Santander-Syro \etal~showed that in conducting \STO~ surface the degeneracy of the titanium (Ti) t$_{2g}$ bands is removed by the surface, and they split into a lower energy d$_{xy}$ band and d$_{yz}$, d$_{xz}$ bands that are filled at higher gate biases \cite{santander2011two}. Joshua \etal~ introduced an atomic spin-orbit coupling term that mixes the three bands into a new, more complex, band structure \cite{joshua2012universal}. Nernst effect and magnetotransport measurements \cite{lerer2011low} suggested that the mobile band is the one responsible for superconductivity. Another view is by Joshua \etal~, who suggested that a second band becomes populated exactly at a critical density where the superconducting critical temperature T$_\textrm{c}$ is maximal \cite{joshua2012universal}. This issue is still a matter of debate.

In this paper we use the Shubnikov-de Haas (SdH) effect (probing the area of the Fermi surface), as well as the resistivity and the low field limit of the inverse Hall coefficient ${R_H^{-1}}$ (which is dominated by the density of the mobile charge carriers) to study the band structure of the \STO/\LAO\ interface. We show that the SdH frequency and ${R_H^{-1}}$ follow a nonmonotonic behavior similar to the dependence of the superconducting critical temperature T$_\textrm{c}$ and magnetic field H$_\textrm{c}$ on gate bias. Both the SdH frequency and ${R_H^{-1}}$ exhibit an anomalous decrease upon increasing the gate bias beyond maximum T$_\textrm{c}$. We interpret this unconventional decrease as arising from electronic correlations between the Ti t$_{2g}$ bands that are mixed by the atomic spin-orbit interaction. Our calculations show that in this case the population of the mobile band and its density of states are nonmonotonic functions of the chemical potential ($\mu$). This can explain the peculiar gate dependence of the transport properties, as well as the decrease in T$_\textrm{c}$ on the overdoped regime in the (V$_\textrm{g}$, T) phase diagram.

\section {Results}
\subsection{Experimental Data}
We use three samples of $\{100\}$ \LAO/\STO~ interface patterned into Hall bars (see Methods for full description). Sample A was used to study the superconducting dome and the low field Hall effect and Samples B and C were used for studying the low field Hall effect and the gate dependence of the SdH frequency.

\begin{figure}
 \begin{center}
\includegraphics[width=1\hsize]{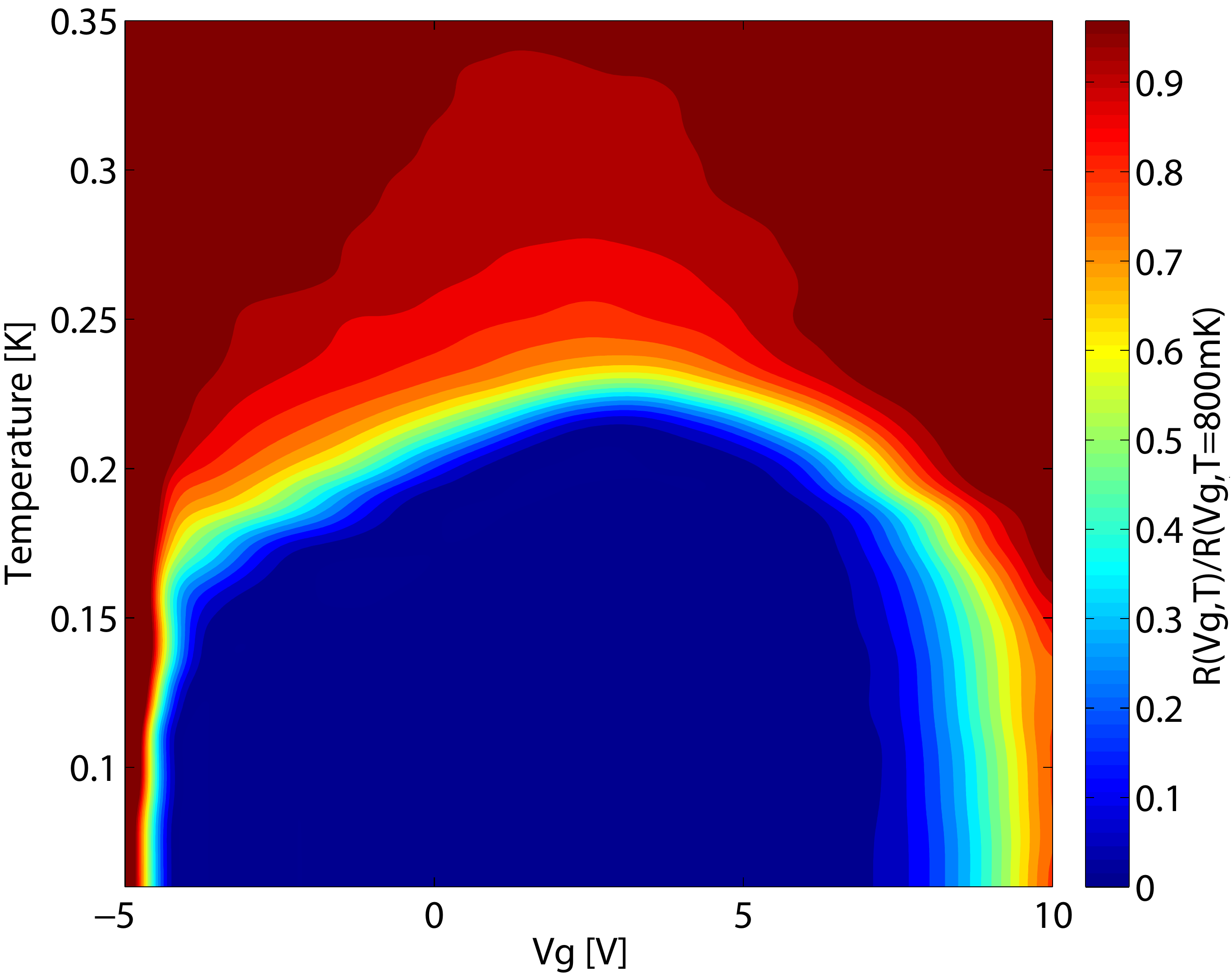}
\caption {Superconducting phase diagram. A color map of the normalized resistance plotted as a function of gate bias and temperature. Each resistance value is divided by the resistance at T=800 mK (above T$_\textrm{c}$) and the same value of V$_\textrm{g}$ (Sample A). \label{phasediagram}}
\end{center}
\end{figure}

Figure~\ref{phasediagram} presents the sheet resistance \Rs\ of Sample A, normalized by its value at T=800 mK for the same gate bias. Defining T$_\textrm{c}$ as the temperature for which R(T$_\textrm{c}$) = R(T=800 mK)/2, the green contour represents T$_\textrm{c}$ as a function of V$_\textrm{g}$. T$_\textrm{c}$ is maximal for V$_\textrm{g}=2.6\pm0.3$V. The entire superconducting dome is revealed by varying V$_\textrm{g}$ by merely 15 V, thus demonstrating the performance of our gate design \cite{rakhmilevitch2013anomalous}.

The shape of the superconducting dome is somewhat different from the previously reported one~\cite{caviglia2008electric}. In particular, in the vicinity of the base of the dome the gate dependence of T$_\textrm{c}$ is very sharp. At both edges of the superconducting dome the equi-resistance lines are almost perpendicular to the gate bias axes. For V$_\textrm{g}\leq-4.8$ V this behavior persists into the normal state and hence may not reflect intrinsic properties of the material, but may arise from the contacts becoming non-ohmic. This region is not relevant for our experiment. The fine-tunability of our devices allows us to study in detail the dependence of the magneto-transport properties and the superconductivity on gate bias.

\begin{figure}
 \begin{center}
\subfigure{\includegraphics[width=1\hsize]{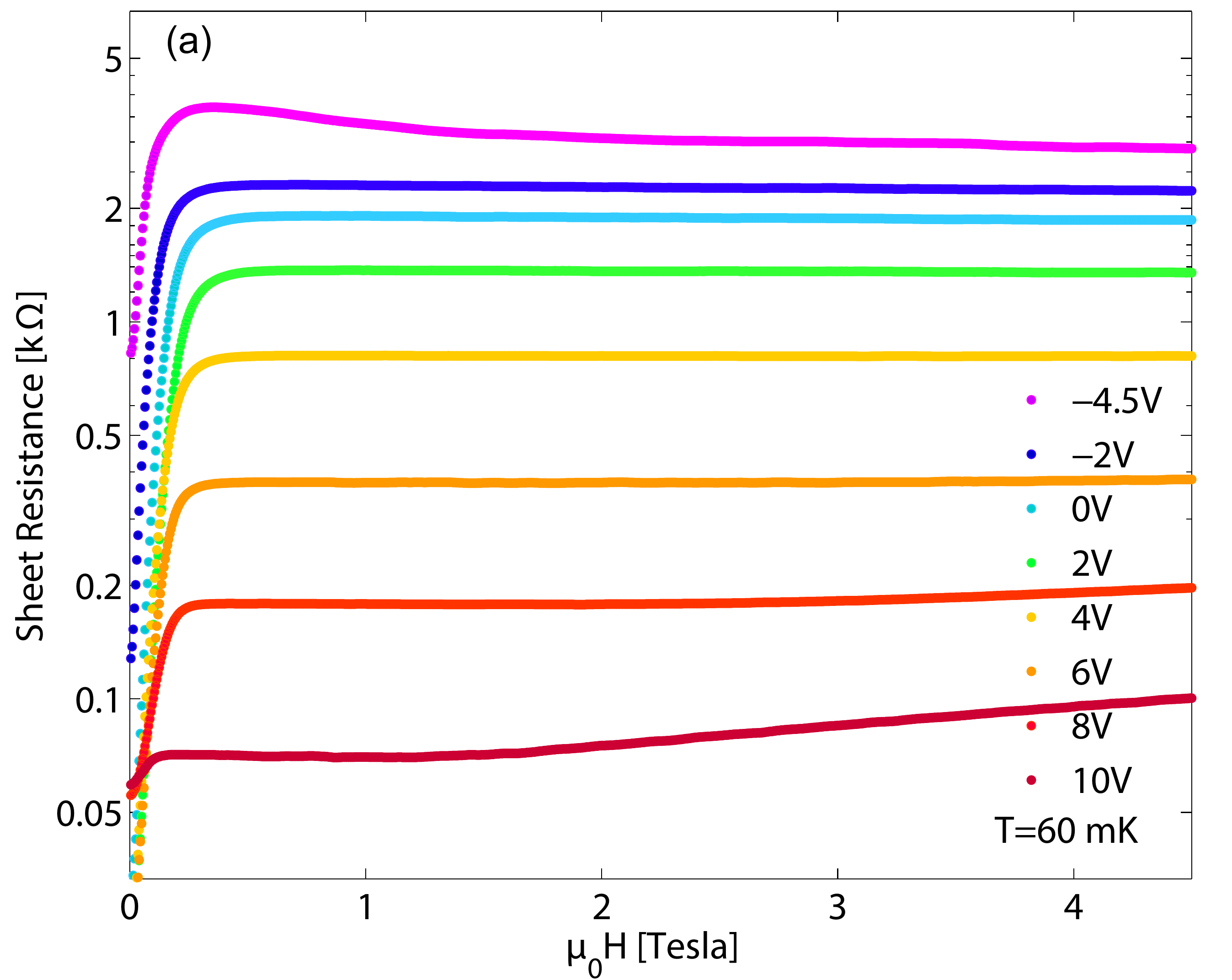}}
\subfigure{\includegraphics[width=1\hsize]{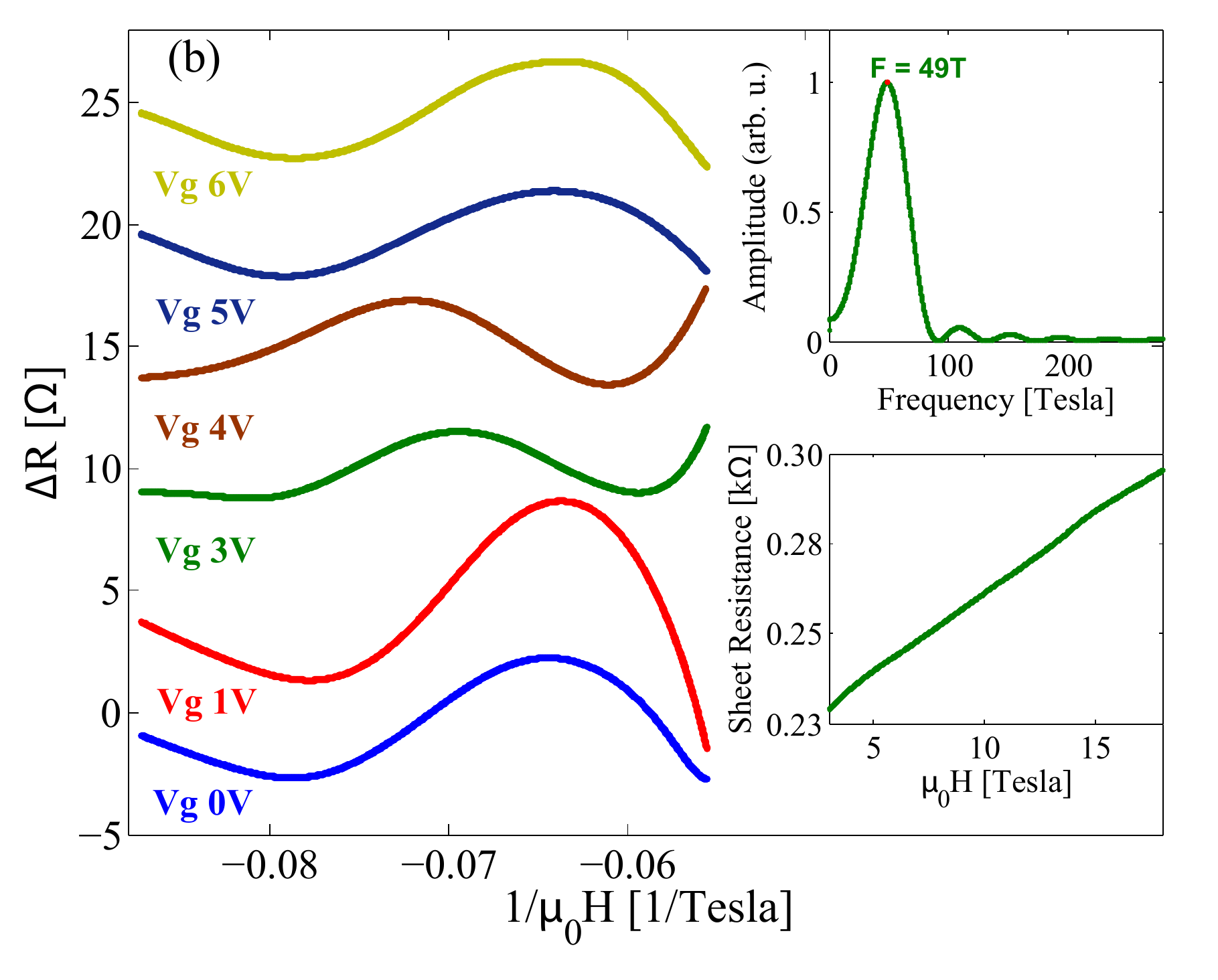}}
\caption {Low and High field measurements. (a) Sheet resistance is plotted on a logarithmic scale as a function of magnetic field at T=60 mK for various fixed values of gate bias V$_\textrm{g}$ (Sample A). (b) Resistance versus inverse magnetic field after subtraction of a smooth polynomial background for various fixed values of gate bias V$_\textrm{g}$ (Sample B). Successive curves are shifted by 5$\Omega$ for clarity. Lower Inset: The resistance per square at V$_\textrm{g}$=3V versus the magnetic field. Upper Inset: Fast Fourier Transform (FFT) of the V$_\textrm{g}$=3V data.} \label{highfield}
\end{center}
\end{figure}

Figure~\ref{highfield}(a) presents the longitudinal resistivity as function of magnetic field for several fixed gate biases for Sample A. Superconductivity is suppressed at fields greater than the critical magnetic field, H$_\textrm{c}$, defined as R(H$_\textrm{c}$)=R(T=800mK)/2.

In the lower inset of Figure \ref{highfield}(b) we show the sheet resistance of Sample B versus high magnetic field for V$_\textrm{g}$=3V. SdH oscillations can be seen. In order to make them clearer, we subtracted a monotonic polynomial background from the data and used a low pass filter to remove noise (Figure \ref{highfield}(b)). Additional data for Sample B is presented in Supplementary Figures 1(b) and 4(a), and for Sample C in Supplementary Figures 2 and 4(b). The upper inset shows a fast Fourier transform used to find the SdH frequency for the data in the lower inset. We found the frequency to be independent of the analysis and background subtraction as explained in the Methods section.

We note that Sample B had both back and top gate electrodes. The latter was prepared using an additional e-beam lithography process, which apparently ruined the superconducting transition. However, all other properties of this sample remained similar to those of samples A, C and other samples previously measured.

\begin{figure}
 \begin{center}
\subfigure{\includegraphics[width=1\hsize]{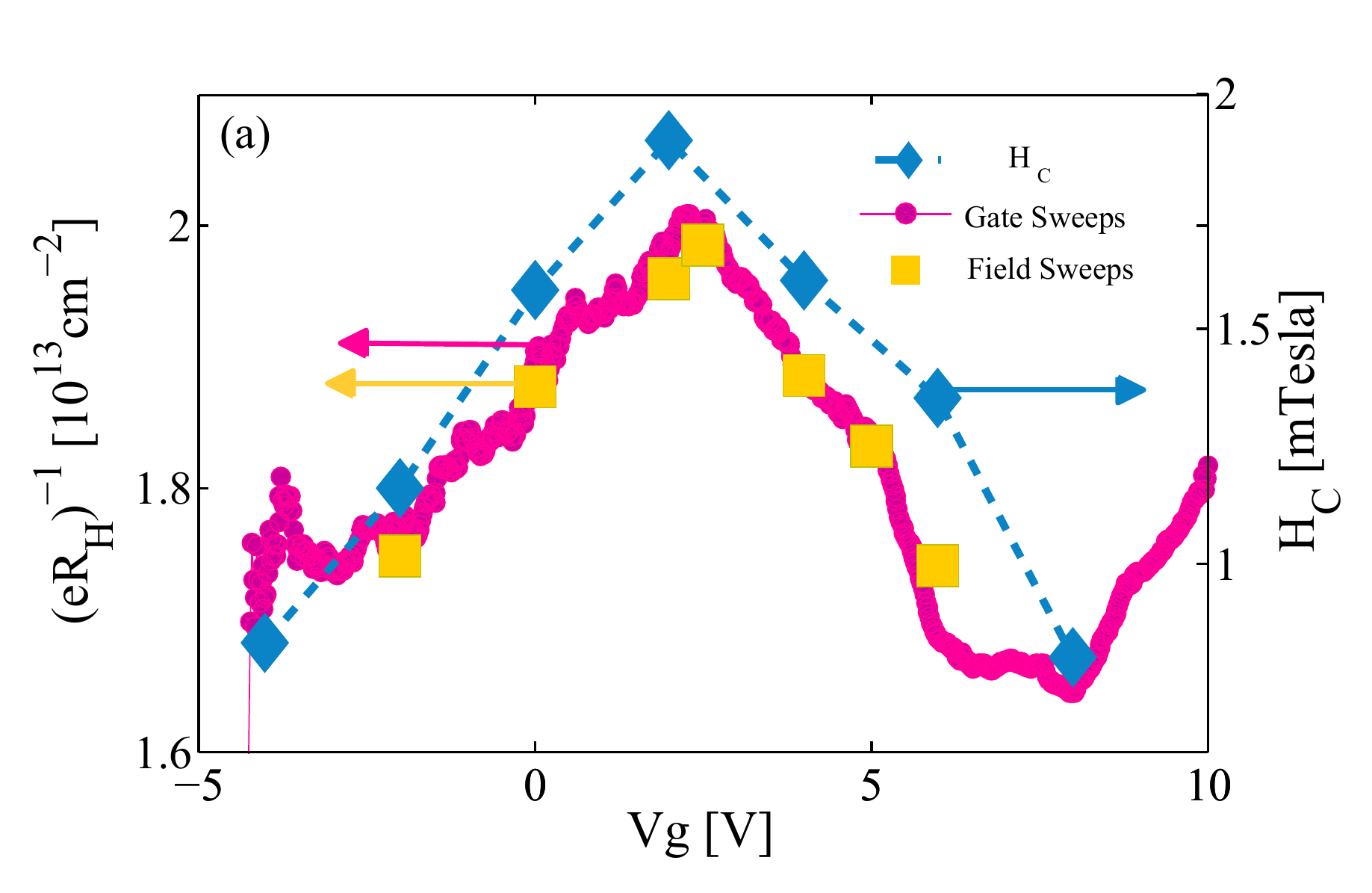}}
\subfigure{\includegraphics[width=1\hsize]{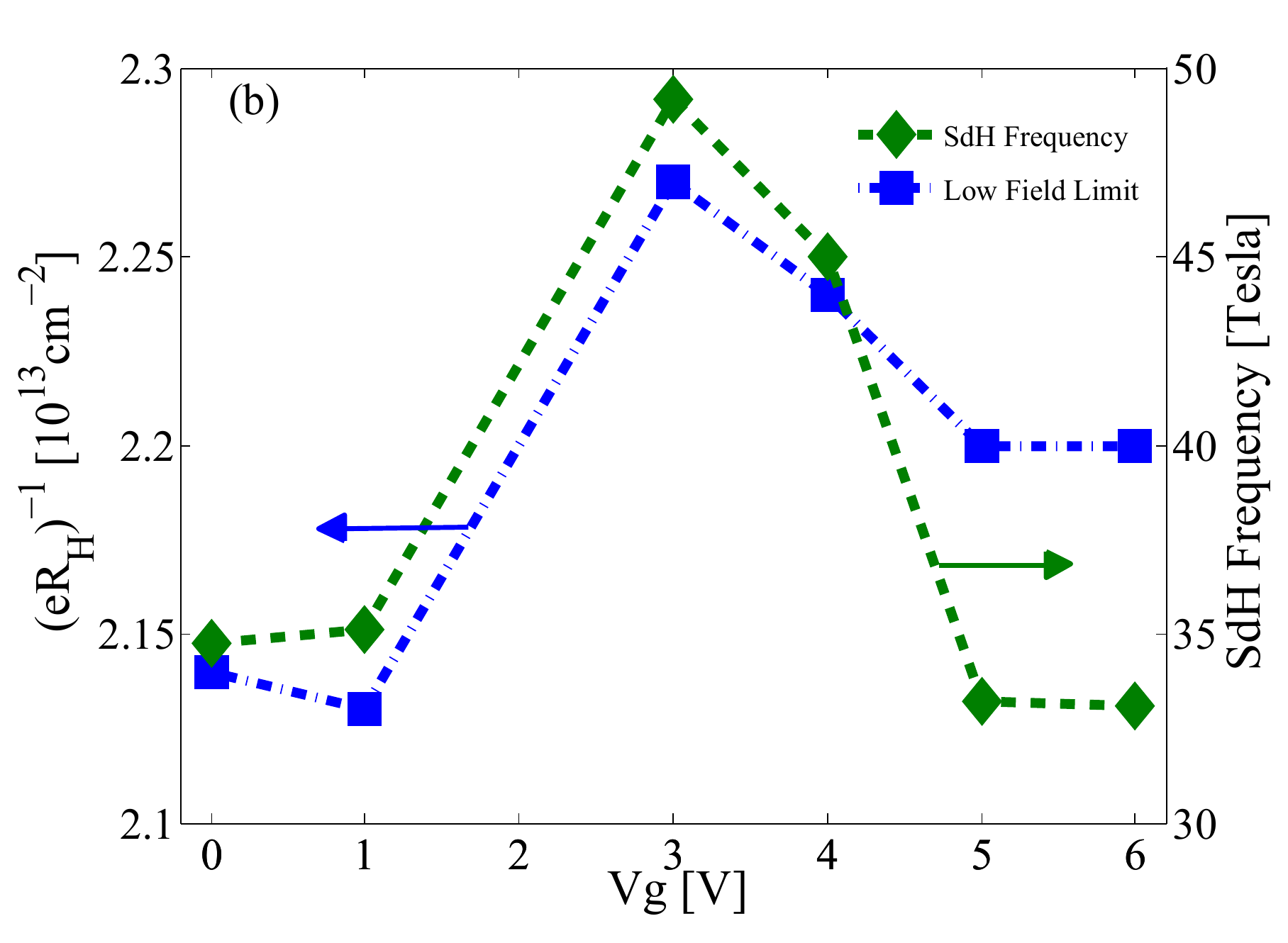}}
\caption {Non-monotonic behavior of the transport and superconducting properties. (a) (Sample A) Left axis (indicated by yellow and magenta arrows): The inverse of the Hall coefficient is plotted as a function of the gate bias for a fixed magnetic field of 3 T (magenta circles). The Hall coefficient was also extracted by measuring the Hall resistance as a function of the magnetic field from 0 T to 3 T and using a linear fit (yellow squares). Right axis (indicated by a blue arrow): Superconducting critical magnetic field H$_\textrm{c}$, defined as $R(T=60mK,H_\textrm{c}) = 1/2R(T=800 mK,H=0)$, plotted for different fixed gates bias values at T=60 mK (blue diamonds). (b) (Sample B) Left axis (indicated by a blue arrow): The inverse of the Hall coefficient inferred from a linear fit to the data in Supplementary Figure 1(c) up to 2T is plotted as a function of the gate bias (blue squares). Right axis (indicated by a green arrow): The SdH frequency is plotted as a function of the gate bias (green diamonds). The SdH frequency is calculated from FFT analysis of the data in Figure 2(b).\label{SCproperties}}
\end{center}
\end{figure}

We shall now examine carefully the gate dependence of the superconducting and transport properties in order to demonstrate that the low field limit of the Hall coefficient, the frequency of the SdH effect, T$_\textrm{c}$, and H$_\textrm{c}$ all exhibit similar nonmonotonic behavior as a function of gate bias. In Figure \ref{SCproperties}(a) the inverse low-field Hall coefficient $(eR_H)^{-1}$ in units of carrier density is plotted as a function of V$_\textrm{g}$ measured at 60mK for Sample A. Below 1K $(eR_H)^{-1}$ is practically temperature independent. Yellow squares are data taken with a fixed V$_\textrm{g}$ while scanning the field from -3T to +3T and fitting a linear curve up to 3 T to the antisymmetric part. The low-field data is completed by the magenta circles, which represent the Hall number inferred from the transverse voltage measured for $\mu_0$H = $\pm$3 T after antisymmetrizing the data. The data taken using these two methods coincide, indicating that the Hall signal is approximately linear in field up to 3T, as can be seen in Figure \ref{SCproperties}(a). A clear non-monotonic signal is observed with a maximum at V$_\textrm{g}$=2.45$\pm0.3$ V. We also plot H$_\textrm{c}$ (blue diamonds), defined by R(H$_\textrm{c}$)=R(T=800mK)/2. H$_\textrm{c}$, T$_\textrm{c}$ and $(eR_H)^{-1}$ are all non-monotonic. The maximum of $(eR_H)^{-1}$ and that of T$_\textrm{c}$ and H$_\textrm{c}$ appear at the same gate bias within error: V$_\textrm{g}$=2.6$\pm0.2$ V.

The nonmonotonic behavior of $(eR_H)^{-1}$ appears to be a universal property of the \LAO/\STO~ interface, independent of \LAO~ layer thickness (between 6 to 16 unit cells). It can also be seen in other reports with various film thickness: e.g., \cite{bell2009dominant}(10 unit cells), \cite{joshua2012universal}(6 and 10 unit cells), \cite{001and110Nonmonotonic2015}(10 unit cells). In Figure~\ref{SCproperties}(a) it is presented for 6 unit cells with a much higher resolution. This behavior is reproduced for 16 unit cells in Figure \ref{SCproperties}(b), and for a third sample of 16 unit cells in Supplementary Figure 3.

To better understand which bands govern the low field Hall coefficient $(eR_H)^{-1}$ we studied it together with the SdH frequency, which is a direct clean measurement of the number of carriers in the mobile band. The SdH signal of less mobile bands is strongly suppressed by its exponential dependence on the Dingle scattering time. We plot $(eR_H)^{-1}$ and the SdH frequency for Sample B as function of back gate bias in Figure~\ref{SCproperties}(b). Blue squares are the Hall coefficients $(eR_H)^{-1}$ at low magnetic fields extracted from a linear fit to the Hall resistance data in Supplementary Figure 1(c) between 0T and 2T. Green diamonds are the SdH frequency, calculated from FFT of the data in Figure \ref{highfield}(b). Both quantities follow the same nonmonotonic behavior, despite an order-of-magnitude difference in the carrier densities they correspond to, as previously noted by Ben Shalom \etal \cite{shalom2010shubnikov}. This means that the population of the mobile band decreases with increasing total number of carriers (increasing gate bias). The nonlinear response of \STO~ to electric field \cite{ChristenDielectricSTO1994} cannot account for the nonmonotonic effects that we observe, since the dielectric function is monotonic as function of the electric field, and, moreover, the total carrier density should in any case increase monotonically with the applied gate voltage. Therefore, such a scenario is possible only if electronic interactions are considered, as we now discuss. Sample C showed the same behavior of decreasing SdH and $(eR_H)^{-1}$ for increasing gate bias (see Supplementary Figure 3).

\subsection{Theoretical Analysis and Results}
To interpret our results we employ a minimal theoretical model of three t$_{2g}$ bands (d$_{xy}$, d$_{xz}$, and d$_{yz}$), split and mixed by atomic spin-orbit coupling, as suggested by Joshua \etal~\cite{joshua2012universal}. The main new ingredient in our model is a Hubbard-type repulsion between electrons occupying different orbitals in the same unit cell. We use a mean-field approximation in which an electron in a certain orbital $i=xy,xz,yz$ experiences an average increase of $U_{ij} N_j$ in its energy due to repulsion by electrons in orbital $j$, with $N_j$ being the average occupancy of electrons in the $j^{th}$ orbital per unit cell (summed over spin). The intra-band interaction between electrons of opposite spins is $\frac{1}{2}U_{ii} N_i$. The orbital populations are calculated self consistently, as explained in the theoretical methods section. For simplicity we took the interaction energy $U_{ij}=U=2.7$eV both for the intra- and inter-band interactions for all orbitals, consistent with Ref.~\cite{PhysRevB.81.153414}.

\begin{figure*}
 \begin{center}
\includegraphics[width=1\hsize]{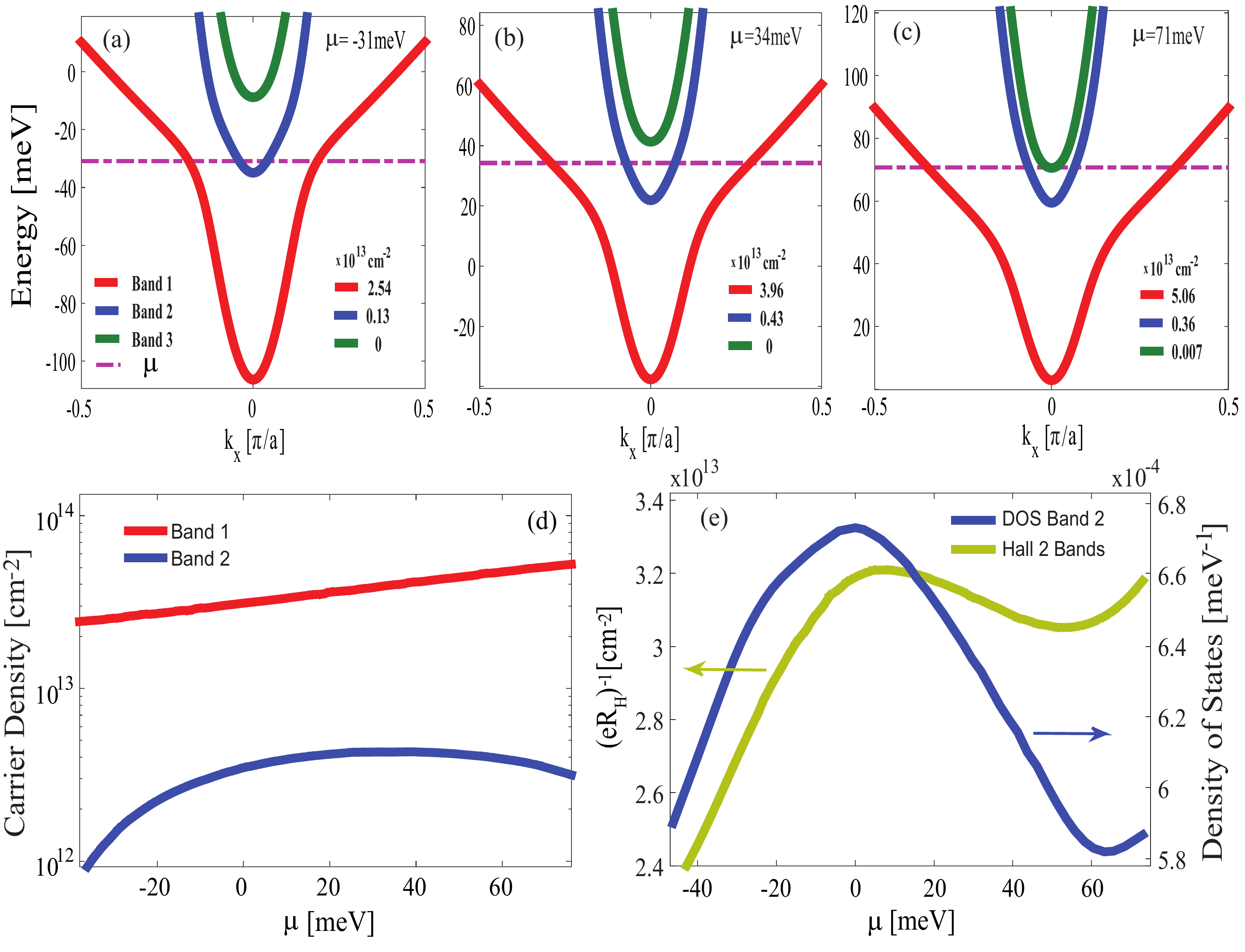}
  \caption{Results of the theoretical model. (a-c) Band structure including the effects of mean-field electronic correlations for three different chemical potentials $\mu$. A cut through the center of the Brillouin zone along the k$_\textrm{x}$ axis is shown. The calculated carrier densities for each band are shown. (d) The carrier concentration of the two lower bands (band 1 and band 2). A non-monotonic behavior of the carrier density of band 2 is obtained. (e) Left axis (indicated by a yellow arrow): Calculated inverse Hall coefficient in the low field limit as a function of $\mu$. Right axis (indicated by a blue arrow): The density of states per unit cell of band 2. Both properties show nonmonotonic behavior with a maximum at a similar value of $\mu$.}\label{Theory}
\end{center}
\end{figure*}

Due to the interactions, upon increasing the chemical potential $\mu$ (i.e., increasing the gate bias) the band structure is modified, as demonstrated in Figure~\ref{Theory}(a)--(c), where the band structure is presented for different values of $\mu$. Each of the three bands should be slightly split due to the Rashba spin-orbit coupling. We neglect this effect since it is very small, as demonstrated in Supplementary Figure 6 (although it is important for the parallel-field dependence of superconductivity \cite{shalom2010tuning} and for the weak anti-localization effect \cite{caviglia2010tunable}).

In addition, scanning SQUID measurements detected free magnetic moments at the interface \cite{bert2011direct} and magnetization measurement \cite{li2011coexistence, AriandoSquid} found evidence for a magnetic phase at elevated temperatures. However, the latter measurements were performed at significant magnetic fields, presumably sufficient to align the moments. We have recently found evidence for a magnetic order at zero field below 1K \cite{AlonRonMagnetic}. The energy scale of this interaction is very low compared with the terms in the presented Hamiltonian. Therefore, and in order to keep our calculations simple, we did not allow for a Stoner instability.

The resulting total densities of electrons in each band are presented in Figure \ref{Theory}(d). While the electron density in the lowest-energy band 1 [red line in Figure \ref{Theory}(a)--(c)] is monotonically increasing with $\mu$, the electron density in band 2 which is populated next [blue line in Figure \ref{Theory}(a)--(c)] has a nonmonotonic dependence on $\mu$ (note that the overall electron density is still monotonic with $\mu$). In Figure \ref{Theory}(e) we present the calculated single-particle density of states (DOS) for band 2, which exhibits a similar nonmonotonic $\mu$-dependence.

To calculate the Hall resistance we employ a model of point impurities. In the absence of orbital mixing this would lead to momentum-independent scattering rates for each band.
In that case, in the two-band regime, the low-field limit of the inverse Hall coefficient $(eR_H)^{-1}$ would be
$(eR_{H})^{-1} = \frac{(n_{1} \zeta_{1}+n_{2} \zeta_{2})^{2}}{n_{1}\zeta_{1}^{2}+n_{2}\zeta_{2}^{2}}$,
with $\zeta_i$ being the mobility and $n_i$ the carrier density of the unmixed $i$th band (orbital).
With band mixing the situation is more complicated, since the composition of the bands in terms of the original t$_{2g}$ orbitals changes as function of the momentum and as function of $\mu$.
We performed the corresponding calculations, taking the carrier densities from Figure \ref{Theory}(d), and assigning mobilities of 6000 and 500 cm$^2/$V$\cdot$s to the unmixed d$_{xy}$ and d$_{yz}$/d$_{xz}$ bands, respectively, roughly representing the ratio of the unmixed bands' effective masses.
The resulting low-field inverse Hall coefficient $(eR_H)^{-1}$ is shown in Figure \ref{Theory}(e). The calculated $(eR_H)^{-1}$ is nonmonotonic as function of the chemical potential, and its maximum appears at the same $\mu$ as the maximum of the DOS of band 2 (blue). This behavior is similar to the observed dependence of the measured $(eR_H)^{-1}$ as function of gate bias for all samples.

\section{Discussion}
In a simple capacitor one expects the carrier concentration to monotonically increase with gate voltage (even in the presence of nonlinear dielectric materials such as \STO). By contrast, in our experiment we see a nonmonotonic behavior of the SdH frequency and $(eR_H)^{-1}$ as function of gate bias. Our theoretical model does give rise to such nonmonotonic behavior as seen in Figure \ref{Theory}.
What is the reason for this theoretical result?
Electronic interactions cause a competition between the occupancies of different bands: When the population of one band increases, the energies of the states in the other band are shifted and the other band's occupation decreases. In the presence of such a competition occupying the band with the highest single-particle density of states is usually preferable, since this results in lower single-particle energy. As Figure~\ref{Theory}(a)--(c) reveals, at low energies band 1 is mainly composed of low effective mass (low density of states) d$_{xy}$ orbitals, but at higher energies it gains a heavier (high density of states) d$_{xz}$/d$_{yz}$ character; the opposite occurs for the band 2.
Thus, as the gate bias, and hence $\mu$, are increased from large negative values, the following occurs: (i) Initially band 1 is occupied, since its energy is lower. As it becomes progressively filled, the energy of band 2 increases [Figure~\ref{Theory}(a)]; (ii) At some point $\mu$ reaches the bottom of the shifted band 2, where band 2's density of states is large, and it begins to fill up [Figure~\ref{Theory}(b)]; (iii) At higher $\mu$, when band 1 becomes heavier than band 2, it becomes preferable to increase the population of band 1 further at the expense of band 2 [Figure~\ref{Theory}(c)]. The population of band 2 is thus nonmontonic as function of $\mu$. A similar population switching effect has recently been discussed in the context of quantum dot physics, where a narrow (high density of states) level  may increase in occupancy at the expense of a  broad (low density of states) one (See, e.g.,\cite{field93,lindemann02,johnson04,baltin99,silvestrov00,goldstein09}).

\begin{figure}
 \begin{center}
\subfigure{\includegraphics[width=1\hsize]{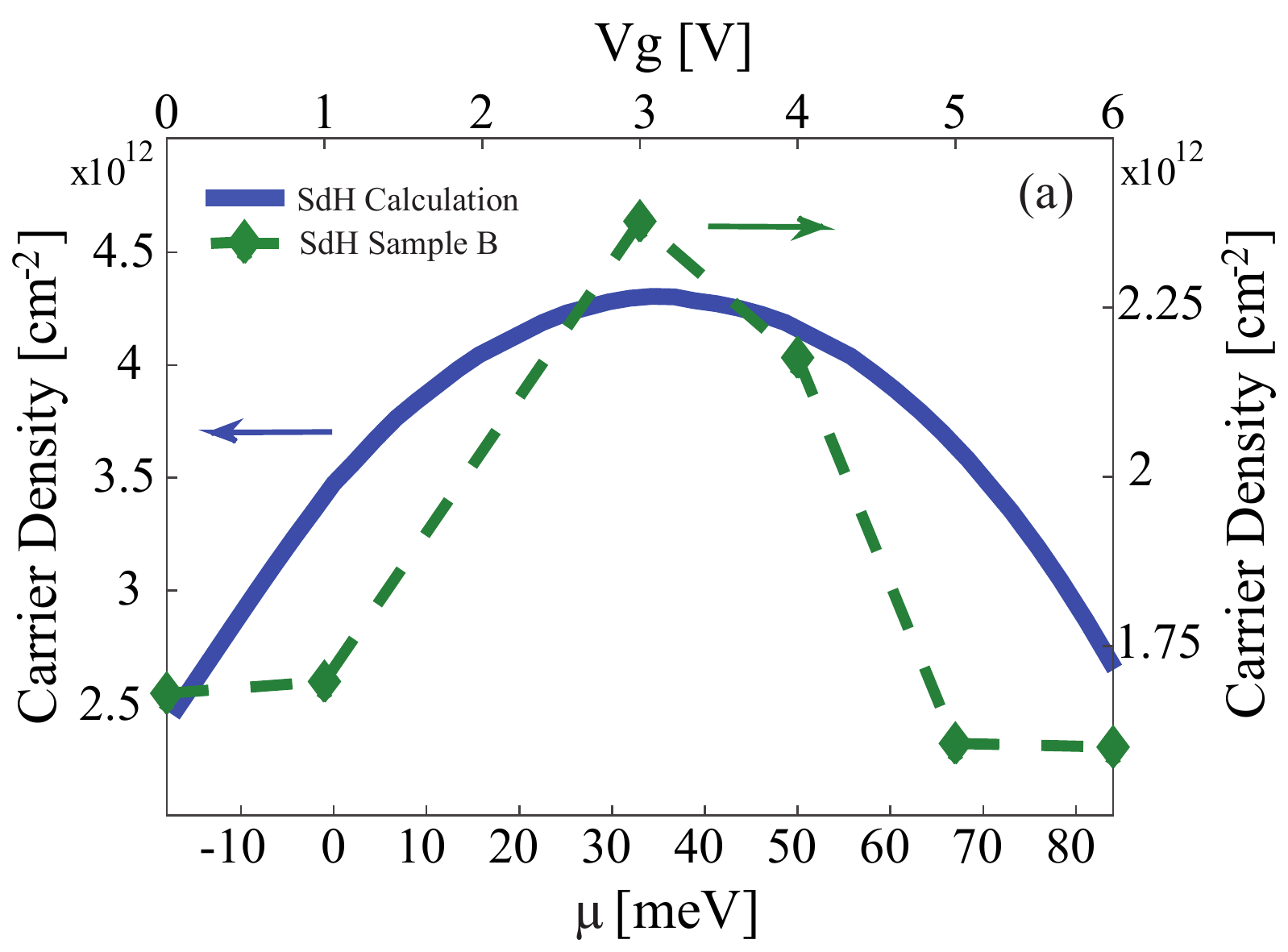}}
\subfigure{\includegraphics[width=1\hsize]{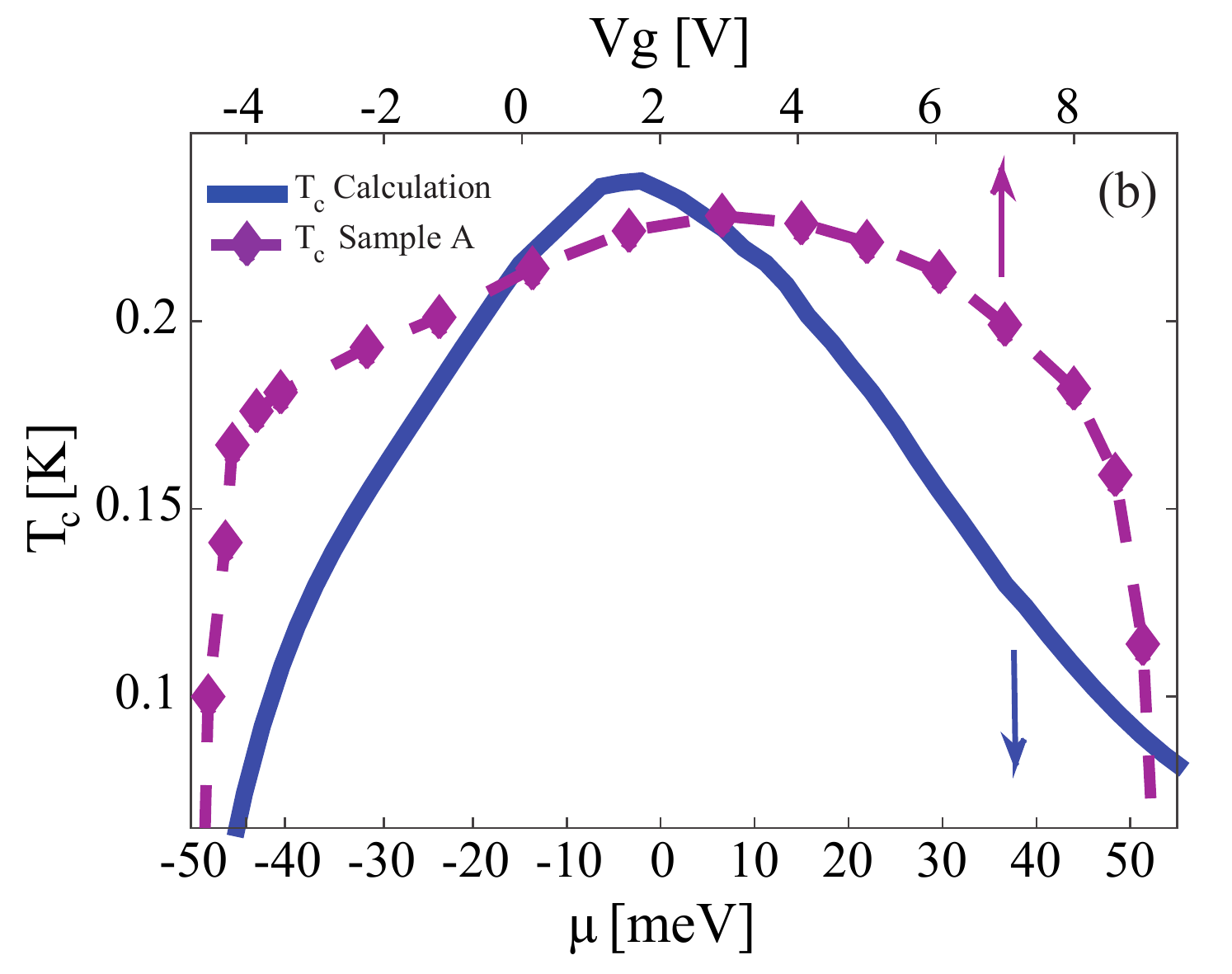}}
  \caption{Comparing experimental data and theoretical calculations. (a) Bottom-Left axes (indicated by a blue arrow): The carrier density of band 2 [taken from Figure 4(d)] is plotted as a function of $\mu$. Top-Right axes (indicated by a green arrow): The carrier density inferred from the SdH frequency in Figure 3(b) is plotted as a function of the gate bias V$_\textrm{g}$. Both show a similar behavior as function of V$_\textrm{g}$ or $\mu$ (which are related through the gate capacitance). (b) Top axis (indicated by a purple arrow): Critical temperature of Sample A (taken from Figure 1) is plotted as a function of V$_\textrm{g}$. Bottom axis (indicated by a blue arrow): Critical temperature calculated from the weak-coupling BCS formula using the density of states of band 2 [taken from Figure 4(e)] is plotted as a function of chemical potential $\mu$. Both critical temperatures exhibit a nonmonotonic behavior with a maximum at the same $\mu$ (V$_\textrm{g}$) as the inverse Hall coefficient.}\label{TheoryVsExp}
 \end{center}
\end{figure}

In Figure \ref{TheoryVsExp}(a) we compare the calculated carrier density of band 2 (blue) and the one extracted from the frequency $F$ of the SdH signal for Sample B through $n=eF/\pi\hbar$. First, we note that the calculation gives the right order of magnitude for the carrier density. Second, the nonmonotonic behavior of the SdH frequency is reproduced by the calculation. The chemical potential was shifted and scaled such that the maximum in the calculated carrier density and the maximal measured SdH frequency coincide. This scaling is consistent with the estimated capacitance of our devices \cite{rakhmilevitch2013anomalous}. Furthermore, Since both bands contribute to the Hall coefficient, while only the mobile second band (whose population is suppressed by the interaction effects) contributes to the SdH, our model naturally explains the order of magnitude difference between $(eR_H)^{-1}$ and the density corresponding to the SdH frequency, which has hitherto been a puzzle \cite{shalom2010shubnikov}.

Our analysis implies that band 1 has much lower mobility than band 2. This could be due to the fact that in the gate voltage region corresponding to the nonmonotonic behaviors band 1 is mainly composed of the heavier d$_{xz}$-d$_{yz}$ orbitals, whereas band 2 mainly includes the lighter d$_{xy}$ orbital. It should be noted that at lower values of $\mu$ (before band 2 enters) band 1 has a d$_{xy}$ character, but is still not observed in SdH. This can be accounted for using the density functional theory calculations of Delugas \etal \cite{DelugasDistancefromInterface}. They find that at low values of $\mu$ the d$_{xy}$ orbital resides in Ti atoms at the STO/LAO interface, where scattering centers are expected to reside, while at higher values of $\mu$ the d$_{xy}$ orbital moves to Ti atoms away from the interface. This picture is also in line with the transport measurements of Bell \etal \cite{bell2009dominant}, and with the observed difference between the effects of top and bottom gates on the transport properties \cite{hosoda2013transistor}. These two gate configurations can effect the location of the wave function along the z direction in an opposite way.

Since the theoretically-calculated DOS $\rho_2$ of the band 2 (blue) has a similar nonmonotonic behavior as the measured $(eR_H)^{-1}$ and T$_\textrm{c}$, it is tempting to try and compare the measured T$_\textrm{c}$ with the weak-coupling single-band BCS formula, T$_\textrm{c} \approx 1.13$T$_\textrm{D} e^{-1/\rho_2 V_\mathrm{BCS}}$. In order to obtain the correct order of magnitude for T$_\textrm{c}$ with the measured Debye temperature T$_\textrm{D}= 400 K$ of \STO\ \cite{Burns1980-SpecificHeat}, we use electron-phonon coupling energy $V_\mathrm{BCS}=0.196$~eV. This value is within the weak coupling limit for all the calculated DOS values used.

In Figure~\ref{TheoryVsExp}(b) we plot the calculated T$_\textrm{c}$ together with the measured one (taken from Figure \ref{phasediagram}) with $\mu$ and V$_\textrm{g}$ scaled to match the superconducting dome region. The nonmonotinc behavior of T$_\textrm{c}$ is reproduced. Using the same scaling the calculated and measured $(eR_H)^{-1}$ showed the same behavior as a function of $\mu$ and V$_\textrm{g}$ (see Supplementary Figure 5). While using a single-band BCS model is probably an oversimplification (and indeed for the low carrier densities a non-BCS type behavior has been reported in planar tunneling measurements \cite{mannharttunneling}), it is interesting to see that it can nicely capture the behavior of T$_\textrm{c}$ versus gate bias in a \STO/\LAO\ interface, and in particular account for the decrease in T$_\textrm{c}$ on the overdoped regime.

In Summary, we found that the Shubnikov-de Haas frequency, the inverse low field Hall coefficient and the superconducting transition temperature all exhibit a similar nonmonotonic dependence on gate bias. In order to explain our data we employed a model involving electronic correlation between the spin-orbit split titanium t$_{2g}$ bands. In this model the second band, which is populated in a non-monotonic fashion at higher gate bias values, is the one responsible for the SdH effect and for superconductivity.

\section {Methods}

\subsection{Experimental}
We grow epitaxial layers of  \LAO~ using reflection high energy electron diffraction (RHEED) monitored pulsed laser deposition on atomically flat TiO$_2$ terminated $\{100\}$ \STO~ 0.5mm thick substrates in standard conditions, oxygen partial pressure of $1\cdot10^{-4}$ Torr and temperature of $780^oC$, as described in \cite{shalom2009anisotropic}. In the first step two unit cells of \LAO~ are deposited. Then Hall bars of $9\times3$ $\mu m^{2}$ (A) $12\times3$ $\mu m^{2}$ (B) and $20\times5$ $\mu m^{2}$ (C) are patterned using electron beam (A) or optical (B and C) lithography followed by deposition of a 40 nm thick amorphous BaTiO$_3$ layer and lift-off to define the conducting channel, where an additional layer of \LAO~ is epitaxially grown in the final step. The total \LAO~ thickness is 6 (16) unit cells for Sample A (B and C). The design aims to minimize screening from the contact pads and leads, thus enhancing the gate bias effectivity. The strong response to back-gating is possible for dielectric substrates with high permittivity (high $\epsilon$) when the distance $d$ between the gate electrode and the tunable surface (or interface) becomes much larger than the width of the mesoscopic conducting channel $\ell$. This results in a capacitance per area of the order of $5\times 10^{12}$ [electrons$\times$cm$^{-2}$Volt$^{-1}$] for $d=0.5$ mm, $\ell$=3 $\mu$m, an order of magnitude larger than for standard planar capacitor geometry \cite{rakhmilevitch2013anomalous}.

Gold gate electrodes are evaporated to cover the back of the substrate. The leakage current is unmeasurably small ($<$1pA). The resistivity and low field measurements of Sample A were performed in a dilution refrigerator with a base temperature of 60 mK using a Princeton Applied Research 124A lock-in amplifier. During cool down, the gate bias was scanned back and forth while recording the temperature and measuring the sheet (per-square) resistance. The gate scan is reversible within a 2 mV resolution, as long as the maximal gate bias (10 V) is not exceeded. The high field measurements were performed on Sample B and C. Sample B was measured in a dilution refrigerator with a base temperature of 20mK at magnetic fields of up to 18 Tesla and a constant top gate voltage of 1 V. Sample C was measured in a He3 cryostat with a base temperature of 350mK and at magnetic fields of up to 34.5 T. Both samples B and C were measured using a Lakeshore 370 resistivity bridge with 3716L low resistance preamplifier and scanner at Tallahassee National High Magnetic Field Laboratory.

For the SdH analysis we first subtracted a smooth polynomial background as described in ref \cite{shalom2010shubnikov}. We made sure that the background consists of no oscillations and that the frequency is independent of background chosen. The background was fitted using a polynomial with a maximum order of 4. The frequency is found both by using FFT analysis with a step-type window function and manually (using the distance between the extrema in the data). Both methods give similar results within an error margin that we estimate to be 5-10\%.

\subsection{Theoretical}
We describe our system using the model introduced in Ref.~\cite{joshua2012universal}, to which we add local (Hubbard) interactions in the mean-field approximation.
In $\mathrm{k}$-space the Hamiltonian is a $6 \times 6$ matrix in the basis of 3 orbitals ($i=xy,xz,yz$) $\otimes$ 2 spin states ($s=\uparrow,\downarrow$). It is a sum of three terms,
$H(\mathrm{k}) = H_0(\mathrm{k})  + H_{SO}(\mathrm{k}) + H_\mathrm{int}(\mathrm{k})$, the first of which being
\begin{widetext}
\begin{equation}
  H_0 (\mathrm{k}) =
  \left(
  \begin{matrix}
     \varepsilon_{xy}(\mathrm{k}) - \Delta_E & 0 & 0 \\
     0 & \varepsilon_{xz}(\mathrm{k}) & 2 t_d \sin(k_x a) \sin(k_y a)\\
     0 & 2 t_d \sin(k_x a) \sin(k_y a) & \varepsilon_{yz}(\mathrm{k}) \\
  \end{matrix}
  \right)
  \otimes I_2
\end{equation}
with
\begin{align}
  \varepsilon_{xy}(\mathrm{k}) = &
  2 t_l \left[ 2 - \cos(k_x a) - \cos(k_y a) \right]
  \\
  \varepsilon_{xz}(\mathrm{k}) = &
  2 t_l \left[ 1 - \cos(k_x a) \right] + 2 t_h \left[ 1- \cos(k_y a) \right]
  \\
  \varepsilon_{yz}(\mathrm{k}) = &
  2 t_h \left[ 1 - \cos(k_x a) \right] + 2 t_l \left[ 1- \cos(k_y a) \right],
\end{align}
\end{widetext}
where $a = 3.905\mathrm{\AA}$ is the \STO\ lattice constant, $t_{l} = 875 meV$, $t_{h} = 40 meV$, and $t_{d} = 40 meV$ are, respectively, the light, heavy, and yz-xz mixing tunneling matrix elements. Through the relation $t_{l/h} = \hbar^2/(2 m_{l/h} a^2)$, these correspond to light/heavy effective masses of $m_{l}=0.7m_{e}$ and $m_{h}=15m_{e}$, respectively. $\Delta_E = 47 meV$ is the relative shift between the $xy$ and $xz/yz$ orbitals due to the confinement potential at the \STO/\LAO\ interface. We neglect orbital coupling due to the asymmetric confining potential and the Rashba term (see Supplementary Figure 6). Following XAS \cite{salluzzo2009orbital}, ARPES \cite{santander2011two}, and transport \cite{joshua2012universal} results, we assume that the other d$_{xy}$ sub-band predicted by DFT calculations \cite{DelugasDistancefromInterface} are localized and do not play a role in transport, following in this respect previous theoretical studies \cite{joshua2012universal,Diez2014Giant}.
\par
Atomic spin orbit coupling contributes
\begin{equation}
  H_{SO}(\mathrm{k}) =
  \frac{i \Delta_{SO}}{2}
  \left(
  \begin{matrix}
    0 & \sigma_x & -\sigma_y \\
    -\sigma_x & 0 & \sigma_z \\
    \sigma_y & -\sigma_z & 0 \\
  \end{matrix}
  \right),
\end{equation}
with $\Delta_{SO} = 40 meV$.
\par
Finally, the mean-field interaction term is
\begin{equation}
  H_{\mathrm{int}} =
  \left(
  \begin{matrix}
    E_{\mathrm{int}}^{xy} & 0 & 0 \\
    0 & E_{\mathrm{int}}^{xz} & 0\\
    0 & 0 & E_{\mathrm{int}}^{yz} \\
  \end{matrix}
  \right) \otimes I_2,
\end{equation}
where $E_{\mathrm{int}}^{i} = \sum_j U_{ij} (1-\delta_{ij}/2) N_j$ 
($i,j=xy,xz,yz$), where $N_j$ is the average occupancy of orbital $j$ (summed over spin) per unit cell. The subtracted term in $E_{\mathrm{int}}^{i}$ removes the unphysical self-interaction of electrons in the same orbital and spin state.
The matrix $U_{ij}$ is symmetric. Moreover, the symmetries of our system dictate that $U_{xy,xz}=U_{xy,yz}$ and $U_{xz,xz} = U_{yz,yz}$. In our calculations we take for simplicity all the element to be equal, $U_{ij} = U = 2.7$eV.
The average occupancies $N_j$ are determined self-consistently, taking into account the contributions of the different eigenstates of the Hamiltonian weighted by their decomposition in terms of the orbitals and the Fermi-Dirac distribution, $f(\varepsilon) = 1/[e^{(\varepsilon-\mu)/k_B T}+1]$:
Denoting the eigenvectors of the matrix $H(\mathrm{k})$ by $\psi_{m,\tau}(\mathrm{k};i,s)$, with corresponding eigenvalues $\varepsilon_{m,\tau}(\mathrm{k})$ ($i=xy,xz,yz$ and $s=\uparrow,\downarrow$ denote the bare orbital and spin, whereas $m=1,2,3$ denote the bands and $\tau = \pm 1$ is a Kramers spin index), we have
\begin{equation}
  N_j = \sum_{m,\tau, s}
  \int_{BZ} \frac{a^2 \mathrm{d}\mathbf{k}}{(2\pi)^2}
  \left| \psi_{m,\tau}(\mathrm{k};j,s) \right|^2
  f \left[ \varepsilon_{m,\tau}(\mathrm{k}) \right],
\end{equation}
where the integration is over the first Brillouin zone.

After the self-consistency equations have been solved, the total density of electrons populating band $m=1,2,3$ (related to the frequency of the SdH oscillations) can be calculated as:
\begin{eqnarray}
  n_m = \sum_{\tau}
    \int_{BZ} \frac{\mathrm{d}\mathbf{k}}{(2\pi)^2}
    f \left[ \varepsilon_{m,\tau}(\mathrm{k}) \right].
\end{eqnarray}
The local density of states (per unit cell) $\rho_m$ of band $m$ at the Fermi energy is the derivative of $a^2 n_m$ w.r.t. $\mu$, keeping the mean-field interaction-induced energy shifts $E^i_{\mathrm{int}}$ constant.

\bibliography{mybib}

\section {Acknowledgements}
We thank I. Neder for useful discussions. Special thanks to J-H Park, D. Graf and Glover E. Jones for help in the magnet lab. This work was supported in part by the Israeli Science Foundation under grant no.569/13 by the Ministry of Science and Technology under contract 3-11875 and by the US-Israel bi-national science foundation (BSF) under grants 2010140 and 2014202. A portion of this work was performed at the National High Magnetic Field Laboratory, which is supported by National Science Foundation Cooperative Agreement No. DMR-0654118, the State of Florida, and the U.S. Department of Energy.

\section {Author contribution}
E.M., M.B.S. and A.R. equally contributed to this work. E.M. A.R, M.B.S and Y.D. designed the experiment, E.M., M.B.S., A.R. and M.M. performed the transport measurements and analyzed the data. E.M., M.B.S. and A.R. prepared the samples. E.M., A.R. and M.G. performed theoretical calculations. All authors discussed the data and wrote the paper.

\section {Additional information}
Supplementary information is available in the online version of the paper. Reprints and permissions information is available online at www.nature.com/reprints. Correspondence and requests for materials should be addressed to Y.D.

\section {Competing financial interests}
The authors declare no competing financial interests.

\end{document}